\documentclass{LMCS}

\def\doi{7 (4:06) 2011}
\lmcsheading%
{\doi}
{1--19}
{}
{}
{Nov.~16, 2010}
{Nov.~24, 2011}
{}

\usepackage{amssymb,graphicx}
\usepackage[american]{babel}
\usepackage{hyphenat}
\usepackage{qtree}
\usepackage{hyperref,enumerate}

\newcommand{\cproblem}[2]{
\noindent
\begin{enumerate}[QUESTIO]
\item[INSTANCE\,:] #1
\item[QUESTION:] #2
\end{enumerate}}

\DeclareMathOperator{\lcp}{lcp}
\newcommand{\yca}{\it{yca}}
\newcommand{\nt}{\,{\nmid}\,}
\newcommand{\cal}{\mathcal}

\begin{document}

\title{The Complexity of Rooted Phylogeny Problems}

\author[M.~Bodirsky]{Manuel Bodirsky\rsuper a}
\address{{\lsuper a}CNRS/LIX, \'Ecole Polytechnique\\Palaiseau, France}
\email{bodirsky@lix.polytechnique.fr}
\thanks{{\lsuper a}The first author has received funding from the European Research Council under the European Community's Seventh Framework Programme (FP7/2007-2013 Grant Agreement no. 257039).}

\author[J.~K.~Mueller]{Jens K. Mueller\rsuper b}
\address{{\lsuper a}Friedrich-Schiller-University\\Jena, Germany}
\email{jkm@informatik.uni-jena.de}

\begin{abstract}
Several computational problems in phylogenetic reconstruction can be formulated
as restrictions of the following general problem: given a formula in
conjunctive normal form where the literals are \emph{rooted triples}, is
there a rooted binary tree that satisfies the formula? If the formulas do not contain disjunctions, the problem becomes the famous rooted
triple consistency problem, which can be solved in polynomial time by an
algorithm of Aho, Sagiv, Szymanski, and Ullman. If the clauses in the formulas
are restricted to disjunctions of negated triples, Ng, Steel, and Wormald
showed that the problem remains NP-complete. We systematically study the
computational complexity of the problem for all such restrictions of the
clauses in the input formula. For certain restricted disjunctions of triples
we present an algorithm that has sub-quadratic running time and is
asymptotically as fast as the fastest known algorithm for the rooted triple
consistency problem. We also show that any restriction of the general rooted
phylogeny problem that does not fall into our tractable class is NP-complete,
using known results about the complexity of Boolean constraint satisfaction
problems. Finally, we present a pebble game argument that shows that the
rooted triple consistency problem (and also all generalizations studied in this
paper) cannot be solved by Datalog.
\end{abstract}

\titlecomment{A preliminary version of this work appeared in the Proceedings of the 13th
International Conference on Database Theory (ICDT), 2010.}

\maketitle

\section{Introduction}

Rooted phylogeny problems are fundamental computational problems for
phylogenetic reconstruction in computational biology, and more generally in
areas dealing with large a\-mounts of data about rooted trees. Given a
collection of \emph{partial information} about a rooted tree, we would like to
know whether there exists a single rooted tree that \emph{explains} the data. A
concrete example of a computational problem in this context is the \emph{rooted
triple consistency problem}. We are given a set $V$ of variables, and a set of
triples $ab|c$ with $a,b,c \in V$, and we would like to know whether there
exists a rooted tree $T$ with leaf set $V$ such that for each of the given
triples $ab|c$ the youngest common ancestor of $a$ and $b$ in this tree is
below the youngest common ancestor of $a$ and $c$ (if such a tree exists, we
say that the instance is satisfiable).

The rooted triple consistency problem has an interesting history.
The first polynomial time algorithm for the problem was discovered
by Aho, Sagiv, Szymanski, and Ullman~\cite{ASSU}, motivated by problems
in database theory. This algorithm was later rediscovered for
phylogenetic analysis~\cite{Steel}. Henzinger, King, and
Warnow~\cite{HenzingerKingWarnow} showed how to use decremental graph
connectivity algorithms to improve the quadratic runtime $ O(mn) $ of the
algorithm by Aho et al. to a deterministic algorithm with runtime $
O(m\sqrt{n})$.

Dekker~\cite{Dekker} asked the question whether 
there is a finite set of `rules' that allows to infer a triple $ab|c$ from
another given set of triples $\Phi$ if all trees satisfying $\Phi$ also satisfy $ab|c$. This question was answered negatively by Bryant and Steel~\cite{BryantSteel}.
Dekker's `rules' have a very natural interpretation in terms of Datalog programs. Datalog as an algorithmic tool for 
rooted phylogeny problems is more powerful than
Dekker's rules. 
We say that a Datalog program \emph{solves}
the rooted triple consistency problem if it derives 
a distinguished 0-ary predicate \emph{false} on a given set of triples
if and only if the instance of the rooted triple consistency problem
is not satisfiable.
One of the results of this paper is the proof that there is no Datalog program that
solves the rooted triple consistency problem.

Datalog inexpressibility results are known
to be very difficult to obtain, and the few existing results often exhibit
interesting combinatorics~\cite{KolaitisVardiDatalog,AfratiCosmadakisYannakakis,FederVardi,GroheThesis,ll}.
The tool we apply to show our result, the existential pebble game, originates in finite model theory, and was successfully applied to finite domain 
constraint satisfaction~\cite{KolaitisVardi}.
A recent generalization of the intimate connection
between Datalog and the existential pebble game to a broad class
of infinite domain constraint satisfaction problems~\cite{BodDalJournal} allows us to
apply the game to study the expressive power of Datalog for the
rooted triple consistency problem.

There are several other important rooted phylogeny problems 
One is the \emph{subtree avoidance problem}, introduced
by~\cite{NgSteelWormald}, or the \emph{forbidden triple problem}~\cite{Bryant};
both are NP-hard. It turns out that all of those problems and many other 
rooted phylogeny problems can be conveniently put into a common framework,
which we introduce in this paper.

A \emph{rooted triple formula} is a formula $\Phi$ in conjunctive normal form where all literals are of the form $ab|c$.
It turns out that
the problems mentioned above and many other rooted phylogeny problems (we provide more examples in Section~\ref{sect:phylo})
can be formalized as the satisfiability problem for a given rooted triple formula
$\Phi$ where the set of clauses that might be used in $\Phi$ is (syntactically) restricted. 
If $\cal C$ is a class of clauses, and the input is confined to rooted
triple formulas with clauses from $\cal C$, we call the corresponding
computational problem the \emph{rooted phylogeny problem for clauses from $\cal C$}.

In this paper, we determine for all classes of clauses $\cal C$ the computational complexity of the rooted phylogeny 
problem for clauses from $\cal C$.
In all cases, the corresponding computational problem is either
in P or NP-complete. In our proof of the complexity classification we apply known results from Boolean constraint satisfaction. 
The rooted phylogeny problem is closely related to
a corresponding \emph{split problem} (defined in Section~\ref{sect:alg}), 
which is a Boolean constraint satisfaction
problem where we are looking for a \emph{surjective} solution,
i.e., a solution where at least one variable is set to true and at
least one variable is set to false. 
The complexity of Boolean split problems has
been classified in~\cite{Creignou}.
If $\cal C$ is such that 
the corresponding split problem can be solved efficiently, our algorithmic results in 
Section~\ref{sect:alg} show that the rooted phylogeny problem for clauses from $\cal C$ can be solved in polynomial time.
Conversely, we present a general reduction that shows that
if the split problem is NP-hard, then the
rooted phylogeny problem for $\cal C$ is NP-hard as well.

\section{Phylogeny Problems}
\label{sect:phylo}
We fix some standard terminology concerning rooted trees.
Let $T$ be a tree (i.e., an undirected, acyclic, and connected graph) with a distinguished vertex $r$, 
the \emph{root} of $T$. The vertices with exactly one neighbor
in $T$ are called \emph{leaves}. The vertices of $T$ are denoted by $V(T)$, and the leaves of $T$ by $L(T) \subseteq V(T)$.
For $u,v \in V(T)$, we say that $u$ \emph{lies below} $v$
if the path from $u$ to $r$ passes through $v$. 
We say that $u$ \emph{lies strictly below} $v$ if
$u$ lies below $v$ and $u \neq v$.
The \emph{youngest common ancestor (yca)} 
of two vertices $u,v \in V(T)$ is the node $w$
such that both $u$ and $v$ lie below $w$ and $w$ has 
maximal distance from $r$. Note that the yca, viewed as a binary
operation, is commutative and associative, and hence there is a canonical definition of the yca of a set of elements $u_1,\dots,u_k$.
The tree $T$ is called \emph{binary} if the root has two neighbors, and every other vertex
has either three neighbors or one neighbor. 
A neighbor $u$ of a vertex $v$ is called a \emph{child} of $v$ (and $v$ is called the \emph{parent} of $u$) in $T$
if the distance of $u$ from the root is strictly larger than the distance of
$v$ from the root.
We write $uv|w$ (or say that $uv|w$ \emph{holds} in $T$) 
if $u,v,w$ are distinct leaves of $T$ and
$\yca(u,v)$ lies strictly below $\yca(u,w)$ in $T$.
Note that for distinct leaves $u,v,w$ of any binary tree $T$, 
exactly one of the
triples $uv|w$, $uw|v$, and $vw|u$ holds in $T$. 

\begin{defi}
A \emph{rooted triple formula} is a (quantifier-free) conjunction of \emph{clauses} (also called \emph{triple clauses}) 
where each clause is a disjunction of 
literals of the form $xy|z$. 
\end{defi}

\begin{exa}\label{ex:nt}
An example of a triple clause is $xz|y \vee yz|x$; it will also be denoted by $xy \nt z$.
Another example of a triple clause is $xy|z_1 \vee xy|z_2$.
\end{exa}

The following notion is used frequently in later sections.
If $\Phi$ is a formula, and $S$ is a subset of the
variables of $\Phi$, then $\Phi[S]$ denotes the conjunction of 
all those clauses in $\Phi$
that only contain variables from $S$.

\begin{defi}\label{def:sat}
A rooted triple formula $\Phi$ 
is \emph{satisfiable} if there exists a rooted binary
tree $T$ and a 
mapping $\alpha$ from the variables of $\Phi$
to the leaves of $T$ such that 
in every clause at least one literal is satisfied.
A literal $xy|z$
is \emph{satisfied} by $(T,\alpha)$
if $\alpha(x),\alpha(y),\alpha(z)$ are distinct and
if
$\yca(\alpha(x),\alpha(y))$ lies 
strictly below $\yca(\alpha(x),\alpha(z))$ in $T$.
The pair $(T,\alpha)$ is then called
a \emph{solution} to $\Phi$.
\end{defi}

We would like to remark that a rooted triple formula
$\Phi$ is satisfiable if and only if there exists a rooted binary
tree $T$ and an \emph{injective}
mapping $\alpha$ from the variables of $\Phi$
to the leaves of $T$ such that the formula evaluates
under $\alpha$ to true.

\begin{exa}
Let $ \Phi = xz|y \vee yz|x \wedge xy|w$ be a rooted triple formula with
variables $ V = \{w,x,z,y\} $. Then the tree $T$
\[
    \Tree [ [ [ x z ] y ] w ]
\]
together with the identity mapping on $V$ is a solution to $\Phi$.
\end{exa}

A fundamental problem in phylogenetic reconstruction is the
rooted triple consistency problem~\cite{HenzingerKingWarnow,BryantSteel,Steel,ASSU}. 
This problem can be stated conveniently in terms of rooted triple
formulas.

\vspace{.2cm}
\begin{prob}[Rooted-Triple-Consistency]
\cproblem{A rooted triple formula $\Phi$ without disjunction.}
         {Is $\Phi$ satisfiable?}
\end{prob}
\vspace{.2cm}

The following NP-complete problem was introduced and studied in an equivalent
formulation by Ng, Steel, and Wor\-mald~\cite{NgSteelWormald}.

\vspace{.2cm}
\begin{prob}[Subtree-Avoidance-Problem]
\cproblem{A rooted triple formula $\Phi$ where each clause is of the form $x_1y_1 \nt z_1 \vee \dots \vee x_ky_k \nt z_k$. (As in Example~\ref{ex:nt},  $xy \nt z$ stands for $xz|y \vee yz|x$.) }
         {Is $\Phi$ satisfiable?}
\end{prob}
\vspace{.2cm}

Also the following problem is NP-hard; it has been studied in~\cite{Bryant}.

\vspace{.2cm}
\begin{prob}[Forbidden-Triple-Consistency]
\cproblem{A rooted triple formula $\Phi$ where each clause is of the form $xy \nt z$.}
         {Is $\Phi$ satisfiable?}
\end{prob}
\vspace{.2cm}

More generally, if $\cal C$ is a class of triple clauses, the 
rooted phylogeny problem for clauses from $\cal C$ is the following computational
problem.

\vspace{.2cm}
\begin{prob}[Rooted-Phylogeny for clauses from $\cal C$]
\cproblem{A rooted triple formula $\Phi$ where each clause can be obtained from clauses in $\cal C$ by substitution of variables.}
         {Is $\Phi$ satisfiable?}
\end{prob}
\vspace{.2cm}

All of these problems belong to NP. A given solution $(T,\alpha)$ can be verified in polynomial time using the following deterministic algorithm. For each literal of each clause of $\Phi$ check whether the literal is satisfied.
If there is at least one literal per clause satisfied by $(T,\alpha)$, then the
given solution is valid else it is invalid. A literal $ab|c$ is satisfied if
$\alpha(a)$, $\alpha(b)$, and $\alpha(c)$ are distinct and if $ v_{1} =
\yca(\alpha(a),\alpha(b))$ lies strictly below $ v_{2} =
\yca(\alpha(a),\alpha(c))$ (recalling definition \ref{def:sat}). Determining the
youngest common ancestor of two vertices is straightforward using a bottom-up search for each vertex.
Another search is then used to check if $v_1$ lies strictly below $v_2$.

Note that the rooted triple consistency problem, the
subtree avoidance problem, and the forbidden triple consistency problem are examples of rooted phylogeny problems, by
appropriately choosing the class $\cal C$. For example, for the rooted triple consistency
problem we choose $\cal C = \{xy|z\}$. The subtree avoidance problem is the rooted phylogeny problem for the class $\cal C$ that 
contains for each $k$ the clause $x_1y_1 \nt z_1 \vee \dots \vee x_k y_k \nt z_k$. 

Finally, note that when $\cal C$ contains clauses with literals of the form $xx|y$,
$xy|x$, or $xy|y$, then these literals can be removed from the clause since they are 
unsatisfiable. 
If \emph{all} literals in a triple clause are of this form, then the clause is unsatisfiable.
It is clear that in instances of the rooted phylogeny problem for clauses from a fixed
class $\cal C$ one can efficiently decide whether the input contains such clauses
(in which case the input is unsatisfiable). Thus, removing such clauses from
$\cal C$ does not affect the complexity of the rooted phylogeny for clauses from $\cal C$.
To prevent dealing with degenerate cases, we therefore make the convention
that all clauses in $\cal C$ do not contain literals of the form $xx|y$,
$xy|x$, or $xy|y$.

\paragraph{\bf Constraint Satisfaction Problems}
Many phy\-lo\-ge\-ny problems can be viewed as infinite domain \emph{constraint
satisfaction problems} (CSPs), which are defined as follows. Let $\Gamma$ be a
structure\footnote{We follow standard terminology in logic, see e.g.~\cite{HodgesLong}.} with a finite relational signature $\tau$. A first-order formula
over $\tau$ is called \emph{primitive positive} if it is of the form $\exists
x_1,\dots,x_n. \, \psi_1 \wedge \dots \wedge \psi_m$ where
$\psi_1,\dots,\psi_m$ are atomic formulas over $\tau$, i.e., of the form $x=y$
or $R(x_1,\dots,x_k)$ for a $k$-ary $R \in \tau$. Then the \emph{constraint
satisfaction problem for $\Gamma$}, denoted by CSP$(\Gamma)$, is the
computational problem to decide whether a given primitive positive sentence
(i.e., a primitive positive formula without free variables) is true in
$\Gamma$. 
The sentence $\Phi$ is also called an \emph{instance} of CSP$(\Gamma)$, and the
clauses of $\Phi$ are also called the \emph{constraints} of $\Phi$.
We cannot give a full introduction to constraint satisfaction and to
constraint satisfaction on infinite domains, but point the reader
to~\cite{CSPSurvey,BodirskySurvey}. Here, we only specify an infinite
structure $\Delta$ that can be used to describe the rooted triple consistency
problem as a constraint satisfaction problem. It will then be straightforward
to see that all rooted phylogeny problems for clauses from a finite class $\mathcal C$ can be formulated as infinite
domain CSPs as well.

The signature of $\Delta$ is $\{|\}$ where $|$ is a ternary relation symbol.
The domain of $\Delta$ is ${\mathbb N} \rightarrow \{0,1\}$, i.e., the set
of all infinite binary strings (hence, the
domain of $\Delta$ is uncountable). 
For two elements $f,g$ of $\Delta$, let
$\lcp(f,g)$ be the set $\{1,\dots,n\}$ where $n$ is the largest natural number 
$i$ such that $f(j)=g(j)$ for all $j \in \{1,\dots,i\}$; 
if no such $ i $ exists, we set $\lcp(f,g) := \emptyset $,
and if $ f = g $, we set $\lcp(f,g) := \mathbb N$. 
The ternary relation
$fg|h$ in $\Delta$ holds on elements $f,g,h$ of $\Delta$ if they are pairwise
distinct and $|\lcp(f,g)| > |\lcp(f,h)|$.

The following lemma shows that instances of the rooted triple consistency problem can be viewed as primitive positive formulas over the signature $\{|\}$.

\begin{prop}
A rooted triple formula $\Phi(x_1,\dots,x_n)$ is satisfiable if and only if the sentence
$\exists x_1,\dots,x_n. $ $\Phi(x_1,\dots,x_n)$ is true in $\Delta$.
\end{prop}
\begin{proof}
Suppose that $\exists x_1,\dots,x_n.$ 
$\Phi(x_1,\dots,x_n)$ 
is true in $\Delta$,
and let $f_1,\dots,f_n: {\mathbb N} \rightarrow \{0,1\}$ be witnesses for
$x_1,\dots,x_n$ that satisfy $\Phi$ in $\Delta$. We define a finite rooted
tree $T$ as follows. The vertex set of $T$ consists of the restrictions of
$f_i$ to $\lcp(f_i,f_j)$ for all $1 \leq i,j \leq n$ (we do not require $i$ and $j$ to be distinct). Vertex $g$
is above vertex $g'$ in $T$ if $g'$ extends $g$; it is clear that this describes $T$
uniquely. Note that  $f_1,\dots,f_n$  are exactly the leaves of $T$,
and that $T$ is binary.
Let $\alpha$ be the map that sends $x_i$ to $f_i$.
Then $(T,\alpha)$ satisfies $\Phi$.

Conversely, let $(T,\alpha)$ be a solution to $\Phi$.  For each vertex $v$ of
$T$ that is not a leaf, let $l(v)$ and $r(v)$ be the two neighbors of $v$ in $T$
that have larger distance from the root than $v$.  Let $h$ be the length of the
path $r=p_1,\dots,p_h=\alpha(x_i)$ from the root $r$ to $\alpha(x_i)$ in $T$.
Define $f_i: \mathbb N \rightarrow \{0,1\}$ by setting $f_i(j) = 0$ if
$p_{j+1}=l(p_j), 1 \leq j < h$, and $f_i(j) = 1$ otherwise.  Clearly, the
elements $f_1,\dots,f_n$ of $\Delta$ show that $\exists x_1,\dots,x_n.
\Phi(x_1,\dots,x_n)$ is true in $\Delta$.
\end{proof}

This shows that the rooted triple consistency problem is
indeed a constraint satisfaction problem. A refined version of this observation will be useful in Section~\ref{sect:datalog} to apply known techniques
for proving Datalog inexpressibility of the rooted triple consistency 
problem.

A triple clause 
is called \emph{trivial} if the clause is satisfied
by any injective mapping from the variables into the leaves of any rooted tree.
The following lemma (Lemma~\ref{lem:simtriple}) shows that the rooted triple
consistency problem is among the simplest rooted phylogeny problems, that is,
for every class $\cal C$ that contains a non-trivial triple clause
the rooted phylogeny
problem for $\cal C$ can simulate the rooted triple consistency problem in a
simple way.

\begin{lem}\label{lem:simtriple}
Let $\phi(x_1,\dots,x_k)$ be a non-trivial triple clause.
Then there are variables $y^1_1,\dots,y^1_k$,
$\dots,y^l_1$,
$\dots,y^l_k \in \{a,b,c\}$
such that $\bigwedge_{i=1..l} \phi(y^i_1,\dots,y^i_k)$ is logically equivalent to $ab|c$.
\end{lem}
\begin{proof}
First observe that if $k=3$ and if $\phi(x_1,x_2,x_3)$ contains only one literal
then renaming its variables is trivial. Second, if $\phi(x_1,x_2,x_3) =
x_1x_3|x_2 \vee x_2x_3|x_1$, then $\phi(a,c,b) \wedge \phi(b,c,a)$ is logically
equivalent to $ab|c$. If $\phi(x_1,x_2,x_3)$ contains three or more
literals, then due to its non-triviality there can only be at most two distinct
literals. Thus, we fall back to one of the already shown cases and the claim
follows for all clauses with exactly three variables.

If $k>3$, then non-triviality of $\phi$ implies that
$\phi(x_1,\dots,x_k)$ can be written as 
$x_{i_1}x_{i_2}|x_{i_3} \vee \phi'(x_1,\dots,x_k)$ for distinct variables $x_{i_1},x_{i_2},x_{i_3}$ such that $\phi'$ does not imply
$x_{i_1}x_{i_2} \nt x_{i_3}$,
or as $x_{i_1}x_{i_2} \nt x_{i_3} \vee \phi'(x_1,\dots,x_k)$ for distinct variables $x_{i_1},x_{i_2},x_{i_3}$ such that $\phi'$
does not imply $x_{i_1}x_{i_2}|x_{i_3}$.
In both cases we can falsify all literals in $\phi'$ that contain a variable $x_{i_4}$ distinct from $x_{i_1},x_{i_2},x_{i_3}$ by making
$x_{i_4}$ equal to some other variable in this literal. The claim then follows from the case $k=3$.
\end{proof}

This implies that the Datalog inexpressibility result for the rooted triple consistency problem we present in the next section applies to all the rooted phylogeny problems for clauses from $\cal C$ that contain a non-trivial clause.

\section{Datalog}
\label{sect:datalog}
Datalog is an important algorithmic concept originating both in logic
programming and in database theory~\cite{AHV,EbbinghausFlum,Immerman}. Feder
and Vardi~\cite{FederVardi} observed that Datalog programs can be used to
formalize efficient constraint propagation algorithms used in Artificial
Intelligence~\cite{Allen,Montanari,Dechter,PathConsistency}. Such algorithms
have also been studied for the phylogenetic reconstruction problem. 
Dekker~\cite{Dekker} studied \emph{rules} that infer
rooted triples from given sets of rooted triples, and asked whether there exists a set of rules 
such that a rooted triple
can be derived by these rules from a set of rooted triples $\Phi$ if and only if it is logically implied by $\Phi$.
This question was answered negatively by Bryant and Steel~\cite{BryantSteel}.

In this section, we show the stronger result that 
the rooted triple consistency problem cannot be solved by Datalog.
This is a considerable strengthening of this previous result by Bryant and Steel, since we can use Datalog programs not only to infer rooted
triples that are implied by other rooted triples, but 
rather might use Datalog rules to infer an arbitrary number of relations 
(aka \emph{IDBs}) of arbitrary arity to solve the problem.
Moreover, we only require that the
Datalog program derives false if and only if the instance is unsatisfiable.
In particular, we do not require that the Datalog program derives every 
rooted triple that is logically implied by the instance (which is required for
the question posed by Dekker). Finally, as already announced in the conference
version of this paper, we show that the proof
technique extends to other constraint formalisms for reasoning about trees. 

In our proof, we use a pebble-game that was introduced 
to describe the expressive power of Datalog~\cite{KolaitisVardiDatalog} and which was later used to study Datalog as a tool for finite domain constraint satisfaction problems~\cite{FederVardi}. The correspondence between Datalog and pebble games extends to infinite domain
constraint satisfaction problems for countably infinite $\omega$-categorical structures. A countably infinite structure is called \emph{$\omega$-categorical} if its first-order theory\footnote{The \emph{first-order theory} of a structure is the set of first-order sentences that are true in the structure.} has exactly one countable model up to isomorphism.
It can be seen (e.g. using the theorem of Ryll-Nardzewski, see~\cite{HodgesLong}) that the structure $\Delta$ introduced in Section~\ref{sect:phylo} is, unfortunately, \emph{not} $\omega$-categorical. 
However, there are several ways of defining an $\omega$-categorical
structure $\Lambda$ (described also in~\cite{Oligo})
which has the same constraint satisfaction problem. 

We exactly follow the axiomatic approach to define such a structure $\Lambda$ given in~\cite{AdelekeNeumann}. A ternary relation $C$ is said to be a \emph{C-relation} on a set $L$ if for all $a,b,c,d \in L$ the following conditions hold:
\begin{enumerate}
\item [(C1)] $C(a; b,c) \rightarrow C(a; c,b)$; 
\item [(C2)] $C(a; b,c) \rightarrow \neg C(b;a,c)$; 
\item [(C3)] $C(a; b,c) \rightarrow C(a; d,c) \vee C(d; b,c)$; 
\item [(C4)] $a \neq b \rightarrow C(a;b,b)$.
\end{enumerate}

\noindent A C-relation is called \emph{dense} if it satisfies
\begin{enumerate}
\item [(C7)] $C(a; b,c) \rightarrow \exists e. \; (C(e;b,c) \wedge C(a; b,e))$.
\end{enumerate}
The structure $(L;C)$ is also called a \emph{C-set}. 

A structure $\Gamma$ is called \emph{$k$-transitive} if for any
two $k$-tuples $(a_1,\dots,a_k)$ and $(b_1,\dots,b_k)$ of distinct elements of $\Gamma$ there is an automorphism\footnote{An automorphism of a structure $\Gamma$ is an isomorphism between $\Gamma$ and itself.} of $\Gamma$ that maps $a_i$ to $b_i$ for all $i \leq k$.
A structure $\Gamma$
is said to be \emph{relatively $k$-transitive} if for every partial isomorphism $f$
between induced substructures of $\Gamma$ of size $k$ 
there exists an automorphism
of $\Gamma$ that extends $f$. Note that a relatively 3-transitive C-set 
is necessarily 2-transitive. 

\begin{thm}[Theorem 14.7 in~\cite{AdelekeNeumann}]
Let $(L;C)$ be a relatively 3-transitive C-set. Then $(L;C)$ is $\omega$-categorical.
\end{thm}

 Theorem 11.2 and 11.3 in~\cite{AdelekeNeumann} show how to construct 
such a C-relation from a \emph{semi-linear order}\footnote{
A poset is connected if for any two $a,b$ there exists a $c$ such that $a \leq c$ and $b \leq c$, or $a \geq c$ and $b \geq c$.
A connected poset is called \emph{semi-linear} if for every point, the set of all points above it is linearly ordered.} 
that is \emph{dense}, \emph{normal}, and \emph{branches everywhere}
(all these concepts are defined in~\cite{AdelekeNeumann}). 
Such a semi-linear order is explicitly constructed in Section 5 of~\cite{AdelekeNeumann}. 

In fact, there is, up to isomorphism, a unique 
relatively 3-transitive countable C-set which 
\begin{enumerate}[$\bullet$]
\item is \emph{uniform with branching number $2$}, that is, if for all $a,b,c \in L$ we have
$C(a;b,c) \vee C(b;c,a) \vee C(c;a,b)$, 
\item is dense, and
\item satisfies $\neg C(a;a,a)$ for all $a \in L$.
\end{enumerate}
(See the comments in~\cite{AdelekeNeumann} after the statement of Theorem 14.7; the condition that $\neg C(a;a,a)$ for all (equivalently, for some)
$a \in L$ has been forgotten there, but is necessary to obtain uniqueness.)

In the following, let $\Lambda$ be the structure whose domain
is the domain of the
dense C-set that is uniform with branching number $2$;
the signature of $\Lambda$ is not the C-relation, but
the relation $xy|z$ defined from the C-relation 
by $$xy|z \; \Leftrightarrow \; C(z;x,y) \wedge x \neq y \wedge y \neq z \wedge x \neq z \; .$$
Structures that are first-order definable in $\omega$-categorical structures are 
$\omega$-categorical (Theorem 7.3.8 in~\cite{HodgesLong}),
so in particular $\Lambda$ is $\omega$-categorical. 
Note that the relation $|$ of $\Lambda$ satisfies (C1), (C2),
(C3), but not (C4). 

The following observation has already been made 
in~\cite{BodirskySurvey}, but without proof, so we provide a proof here.

\begin{prop}
A rooted triple formula $\Phi(x_1,\dots,x_n)$ is satisfiable if and only if the sentence
$\exists x_1,\dots,x_n.$ 
$\Phi(x_1,\dots,x_n)$ is true in $\Lambda$.
\end{prop}
\begin{proof}
Suppose that there are $a_1,\dots,a_n$ 
such that $\Phi(a_1,\dots,a_n)$ 
is true in $\Lambda$. We first define a binary relation $\preceq$
on the set of all pairs $(a,b)$ with
$a,b \in \{a_1,\dots,a_n\}$. 
We set $(a,b) \preceq (c,d)$ if $\neg cd|a \wedge \neg cd|b$, and 
define $R := \{(u,v) \; | \; u \preceq v \wedge v \preceq u\}$.

\begin{lem}[Lemma 12.1 in~\cite{AdelekeNeumann}]
The relation $\preceq$ is a preorder, and hence $R$ is an equivalence relation.
\end{lem}

Also the following is taken from~\cite{AdelekeNeumann}; but to avoid extensive references
into the proofs there, we give a self-contained presentation
here. 
We claim that
the poset $\preceq/R$
that is induced by $\preceq$ in the natural way on the equivalence classes of $R$ is semi-linear. 
To see this, let $(a_1,a_2),(b_1,b_2),(c_1,c_2)$ be such that
$(a_1,a_2) \preceq (b_1,b_2)$ and
$(a_1,a_2) \preceq (c_1,c_2)$. We have to show that
$(b_1,b_2)$ and $(c_1,c_2)$ are comparable in $\preceq$.
If $(b_1,b_2) \not \preceq (c_1,c_2)$, then $c_1c_2|b_1$
or $c_1c_2|b_2$. Suppose in the following that $c_1c_2|b_1$;
the case $c_1c_2|b_2$ is analogous. Since $(a_1,a_2) \preceq (c_1,c_2)$
we have in particular $\neg c_1c_2|a_1$ in $\Lambda$. 
Recall that the relation $|$ satisfies (C3), which 
can be equivalently written as $\forall a,b,c. \; (C(a;b,c) \wedge
\neg C(d;b,c)) \rightarrow C(a;d,c)$, 
so we find that
$a_1c_1|b_1$. 
By (C2) we have $\neg a_1b_1|c_1$. 
Since $(a_1,a_2) \preceq (b_1,b_2)$
we have $\neg b_1b_2|a_1$.
Axiom (C3) can also be written as
$\forall a,b,c. \; (\neg C(a;d,c) \wedge
\neg C(d;b,c)) \rightarrow \neg C(a;b,c)$,
and thus $\neg b_1b_2|c_1$. Similarly,
$\neg b_1b_2|c_2$. Therefore, $(c_1,c_2) \preceq (b_1,b_2)$,
which is what we had to show. 

Next, note that when $(d_1,d_2)$ and $(e_1,e_2)$ 
are incomparable with respect to $\preceq$,
then $(d_1,e_1)$ is an upper bound for
$(d_1,d_2)$ and $(e_1,e_2)$, that is, 
$(d_1,d_2) \preceq (d_1,e_1)$ and $(e_1,e_2) \preceq (d_1,e_1)$.
It follows that $\preceq/R$ is indeed a semi-linear order with a smallest element $r$, and there exists a tree $T$ on the equivalence classes of $R$
such that $p$ lies below $q$ in $T$ if for all (equivalently, for some) $(a,b) \in p$
and $(c,d) \in q$
we have $(c,d) \preceq (a,b)$. 
Let $\alpha$ be the map that sends $x_i$ to the equivalence class of 
$(a_i,a_i)$;
it is straightforward to verify that $(T,\alpha)$ satisfies $\Phi$.

Conversely, let $(T,\alpha)$ be a solution to $\Phi$. 
We now determine elements $a_1,\dots,a_n$ from $\Lambda$,
and prove by induction on $i$ that 
$\alpha(x_r)\alpha(x_s)|\alpha(x_t)$ in $T$ if
and only if $a_ra_s|a_t$ in $\Lambda$, for all $r,s,t \leq i$.
This is trivial for $n=i=1$,
and for $n=i=2$ we can choose arbitrary distinct 
elements $a_1$ and $a_2$ from $\Lambda$. 
Now suppose we have already found elements $a_1,\dots,a_i$ of $\Lambda$, for $2\leq i < n$, that satisfy the inductive hypothesis.
Let $v$ be the vertex in $T$ that
has the maximal distance from the
root of $T$ such that 
 there is an $j \leq i$ 
where both $\alpha(x_j)$ and
$\alpha(x_{i+1})$ lie strictly below $v$. 

First consider the case that $v$ is the root of $T$.
Then we can choose $k,l \in \{1,\dots,i\}$ such that 
$v = \yca(\alpha(x_k),\alpha(x_l))$.
Let $a$ be an element of $\Lambda$ that is distinct from
$a_k$ and $a_l$, and by the properties of $\Lambda$ ($xy|z$ is uniform with
branching number 2)
we have that $a_k a_l | a$, $a_ka|a_l$ or $aa_k|a_l$ holds.
In the first case, we set $a_{i+1}$ to $a$.
In the second case, 
by relative 3-transitivity of $\Lambda$
there exists an automorphism $\beta$ of $\Lambda$ that
maps $a_k$ to $a_l$ and that fixes $a$. In this case
we set $a_{i+1}$ to $\beta(a_l)$.
In the third case we proceed similar to the second.
In all three cases we have $a_p a_q | a_{i+1}$ for all $p,q \leq i$,
which proves the inductive step.

Next, consider the case that $v$ is not the root of $T$.
In this case, there must be an $m \leq i$ such that 
$\alpha(x_j)\alpha(x_{i+1})|\alpha(x_m)$; choose $m$ such that
the distance between the root and $\yca(\alpha(x_{j}),\alpha(x_m))$
is maximal. 
When $j$ is the only index of size at most $i$ such that
$\alpha(x_j)$ lies below $v$ in $T$, then 
density of $\Lambda$ (axiom (C7) in the special case that
$b=c$) implies that there is an
$a$ such that $a_j a | a_m$. We can then set $a_{i+1}$ to $a$.
Otherwise, there are $j',j'' \leq i$ such that 
$\alpha(x_{j'})\alpha(x_{j''})|\alpha(x_{i+1})$; choose
$j', j''$ such that the distance between $v$ and
$\yca(\alpha(x_{j'}),\alpha(x_{j''}))$ is minimal.
Again we apply density (axiom (C7))
and conclude that there is an $a$ such that $a_{j'}a_{j''}|a$ and
$a_{j'}a|a_m$. We can then set $a_{i+1}$ to $a$. 
\end{proof}

\paragraph{\bf The Existential Pebble Game.}
The fact that $\Lambda$ is $\omega$-categorical allows us to use the existential $k$-pebble game to establish the Datalog lower bound for the rooted
triple consistency problem~\cite{BodDalJournal}. 

The existential $k$-pebble game (for a structure $\Gamma)$ is played by the players 
\emph{Spoiler} and \emph{Duplicator} on an instance $\Phi$ of CSP$(\Gamma)$ and $\Gamma$.
Each player has $k$ pebbles, $p_1, \dots, p_k$ for 
Spoiler and $q_1, \dots, q_k$ for Duplicator; we say that that $q_i$ \emph{corresponds} to $p_i$. 
Spoiler places his pebbles on the variables of $\Phi$, 
Duplicator her pebbles on elements of $\Gamma$. 
Initially, none of the pebbles is placed. In each round of the game
Spoiler picks some of his pebbles.
If some of these pebbles are already placed on $\Phi$, 
then Spoiler removes them from $\Phi$, 
and Duplicator responds by removing the corresponding pebbles from $\Gamma$. 
Duplicator looses if at some point of the game 
\begin{enumerate}[$\bullet$]
\item there is a clause $R(x_1,\dots,x_k)$ in $\Phi$ 
such that $x_1,\dots,x_k$ are pebbled by $p_{j_1},\dots,p_{j_k}$, and
\item the corresponding pebbles $q_{j_1},\dots,q_{j_k}$ of Duplicator are placed on
elements $b_1,\dots,b_k$ in $\Gamma$ such that $R(b_1,\dots,b_k)$ does \emph{not} hold in $\Gamma$.
\end{enumerate}
Duplicator wins if the game continues forever. 
We will make use of the following theorem from~\cite{BodDalJournal}.

\begin{thm}[Theorem 5 in~\cite{BodDalJournal}]\label{thm:pebble}
Let $\Gamma$ be an $\omega$-categorical (or finite) structure.
Then there is no Datalog program that solves CSP$(\Gamma)$
if and only if for every $k$ there exists an unsatisfiable instance $\Phi_k$ of
CSP$(\Gamma)$ such that Duplicator wins the existential
$k$-pebble game on $\Phi_k$ and $\Gamma$.
\end{thm}

\paragraph{\bf Our Method.}
The \emph{incidence graph $G(\Phi)$} of an instance $\Phi$ 
of CSP$(\Gamma)$
is the (undirected, simple) bipartite graph whose vertex set is the
disjoint union of the variables of $\Phi$ and the 
clauses of $\Phi$.
An edge joins a variable $a$ and a clause $\phi$ of $\Phi$
when $a$ appears in $\phi$. 
A \emph{leaf} of $\Phi$ is a variable that has degree one 
in $G(\Phi)$. 
An instance has \emph{girth $k$} if the shortest
cycle of its incidence graph has $2k$ edges\footnote{If we view instances in the obvious way as structures rather than formulas, our definition of girth corresponds to the standard definition of girth in graph theory.}.

\begin{lem}\label{lem:dom}
Let $\Gamma$ be an $l$-transitive (for $l \geq 1$) $\omega$-categorical (or finite) structure with relations of arity at most $l+1$. Suppose that for 
every $k$ there exists an unsatisfiable instance $\Phi_k$
of girth at least $k$ where
every constraint has an injective satisfying assignment.
Then CSP$(\Gamma)$ cannot be solved by Datalog.
\end{lem}

We will see examples for $l=1$ and for $l=2$ in this paper.
Note that by $1$-transitivity, every unary relation in $\Gamma$
either denotes the empty set or the full domain of $\Gamma$.
Since $\Phi_k$ only contains satisfiable constraints,
all unary constraints in $\Phi_k$ are satisfied by every mapping to $\Gamma$. So we make in the following the assumption
that $\Phi_k$ does \emph{not} contain unary constraints.

In the proof we use the following concept, 
inspired by a Datalog inexpressibility
result that was established for temporal
reasoning~\cite{ll}.

\begin{defi}
Let $\Phi$ be an instance of girth at least $k+1$.
Then a subset $S$ of at least $2$ and at most $k$ variables of $\Phi$ is called
\emph{dominated} if $G_S := G(\Phi[S])$ is connected (and hence a tree), and if all but at most one of the leaves of
$G_S$ are pebbled.
\end{defi}

The notion of dominated sets allows us to specify a winning strategy for
Duplicator for the existential $k$-pebble game.

\begin{proof}[Proof of Lemma~\ref{lem:dom}]
To apply Theorem~\ref{thm:pebble}, we have to prove that Duplicator
wins the existential
$k$-pebble game on $\Phi_k$ and $\Gamma$.

Suppose that in the course of the game,
 $u$ is an unpebbled leaf of a dominated set $S$ with pebbled
leaves $a_1,\dots,a_l$, and let $b_1,\dots, b_l$ be the corresponding responses
of Duplicator. Duplicator will play in such a way that $b_1,\dots,b_l$ are pairwise distinct. Moreover, Duplicator 
always maintains the following invariant. \emph{Whenever Spoiler
places a pebble on $a_{l+1}$, Duplicator can play a value $b_{l+1}$ from $\Gamma$ such
that the mapping that assigns $a_i$ to $b_i$ for $1 \leq i \leq l+1$ can be extended to all of $S$ such that this extension is a satisfying assignment for $\Phi_k[S]$.} 

The invariant is satisfied at the beginning of the game:
when spoiler places a pebble on $a_1$, Duplicator can
play any value $b_1$, which is a legal move by our assumption
that $\Phi_k$ does not contain unary constraints.

Suppose that
during the game Spoiler pebbles a variable $a$.
Let $S_1,\dots,S_p$ be the dominated sets where $a$ is the unpebbled leaf
before Spoiler puts his pebble on $a$. (If there is no such dominated set, then
$p=0$.) Let $T_1,\dots,T_q$ be the newly created dominated sets after Spoiler
put his pebble on $a$. Note that since each $T_i$ has not been a dominated set
before Spoiler put his pebble on $a$, it must contain one unpebbled leaf distinct from $a$, which we denote by $r_i$. For an illustration,
see Figure~\ref{fig:schema}.

\begin{figure}
\begin{center}
\includegraphics[scale=0.5]{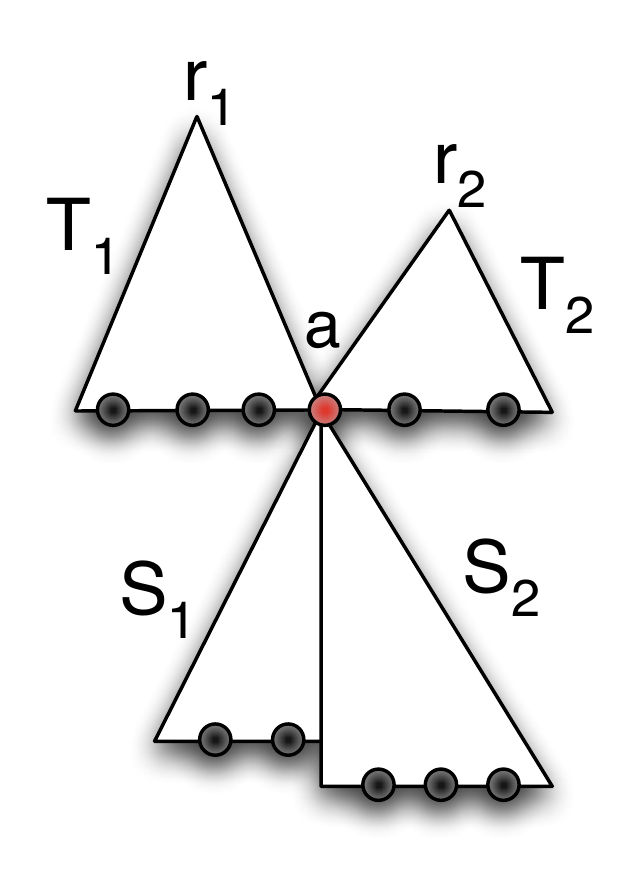}
\end{center}
\caption{A situation in the proof of Lemma~\ref{lem:dom}: Spoiler just pebbled $a$, Duplicator is next.}
\label{fig:schema}
\end{figure}

We have to show that under the assumption that
Duplicator in her previous moves has always maintained the invariant, she will
be able to make a move that again fulfills the invariant. 
If $p>0$, then the union
$S$ of the sets $S_1,\dots,S_p$ was itself a dominated set already before
Spoiler played on $a$, since $G_S$ is clearly connected (all the $S_i$ share the vertex $a$) and no unpebbled leaves can be created by taking a union of dominated sets. The next move of Duplicator is the value $b$
from the invariant applied to $S$.
 This preserves the invariant,
since for every $i \leq q$, the set $T_i \cup S$ has been a dominated set
already before Spoiler played on $a$:
because $T_i$ and $S$ share the vertex $a$, the graph $G_{S \cup T_i}$ is connected, and since $a$ is not a leaf in $G_{S \cup T_i}$, the only unpebbled leaf of $G_{S \cup T_i}$ is $r_i$. 
Therefore, $\alpha$ can be extended to all of $T_i$.

If $p=0$, Duplicator plays an arbitrary
element $b$ in $\Gamma$. 
We prove by induction on the size of $T_i$ that $\alpha$ can be 
extended to $T_i$ such that $\alpha(a)=b$. We can assume that only leaves in $G_{T_i}$ are pebbled (otherwise, since $G_{T_i}$ is a tree, the task reduces to proving the statement for proper subsets of $T_i$).
Consider a clause $\phi$ of $\Phi_k[T_i]$ that contains $a$,
and let $V$ be the variables of $\phi$. This clause must be unique:
otherwise, the graph obtained from $G(\Phi_k)$ by removing the vertex $a$ has
at least two components. 
Only one of those components can contain $r_i$;
the other component must then be a dominated set where all leaves are pebbled,
a contradiction to the assumption that $p=0$. 
Now consider the graph $H$ obtained from $G_{T_i}$ by removing
the vertex that corresponds to $\phi$. 
See Figure~\ref{fig:schema-2}.

\begin{figure}
\begin{center}
\includegraphics[scale=0.5]{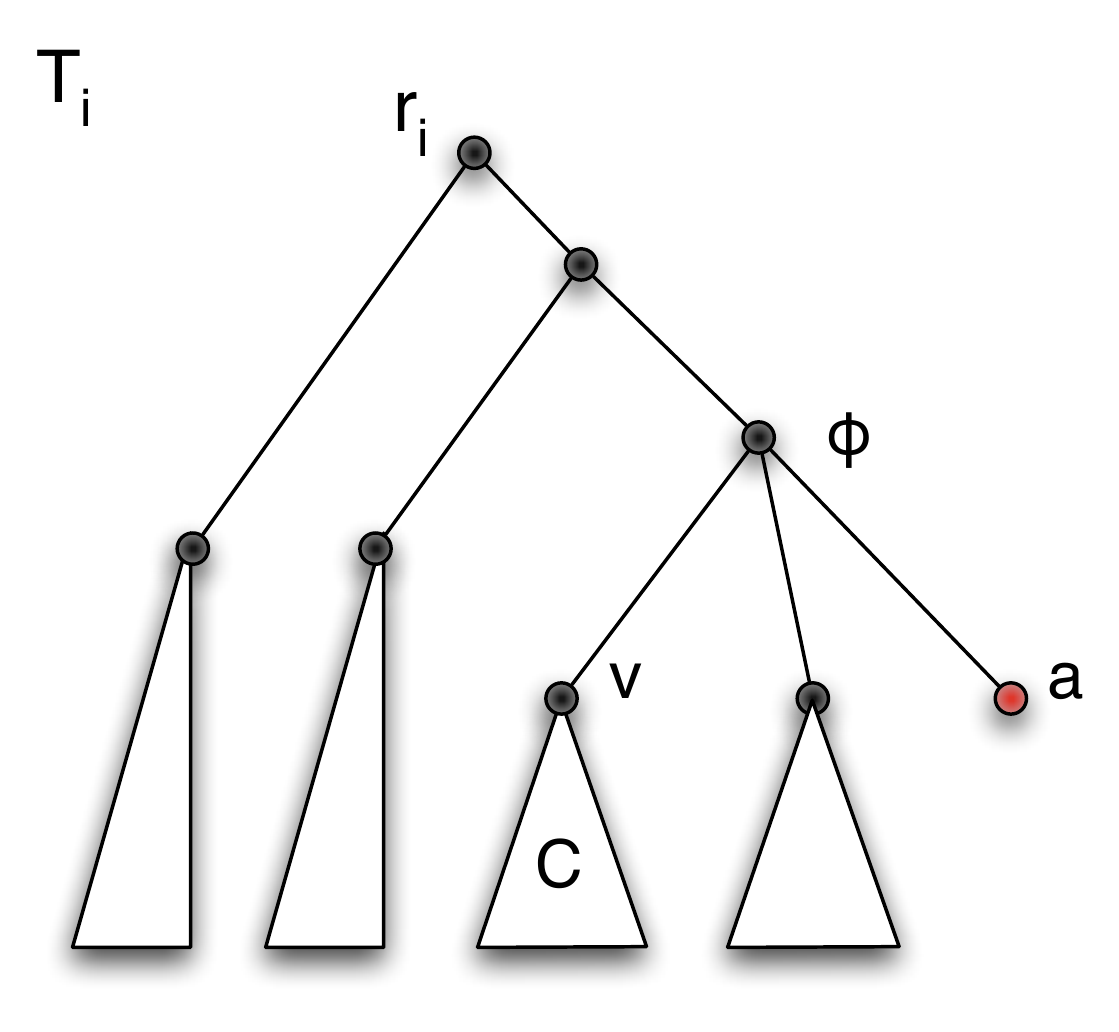}
\end{center}
\caption{A situation in the proof of Lemma~\ref{lem:dom}: extending $\alpha$ to all of $T_i$.}
\label{fig:schema-2}
\end{figure}

If one of the connected components of $H$, say $C$, forms a dominated set, 
then the unique variable $v$ in $C \cap V$ (uniqueness
again follows from the fact that $G_{T_i}$ is a tree) is the unique unpebbled leaf of $C$,
and by the invariant of Duplicator's strategy $\alpha$ can be extended to $\alpha'$ that is defined on all of $C$
such that it satisfies $\Phi_k[C]$. Hence, by removing the pebbles from $C$ and adding a pebble on $v$,
with $\alpha'(v)$ the corresponding response of Duplicator,
we can apply the inductive assumption to $T_i \setminus C \cup \{v\}$ to find an extension of $\alpha$
that is a satisfying assignment for $\Phi_k[T_i]$ and maps $a$ to $b$.

Otherwise, all variables in $V$ except for the variable that lies in the connected component of
$r_i$ in $H$ are pebbled. 
By our assumption on the signature, the clause $\phi$ contains at most
$l$ pebbled variables (including $a$). 
Also by assumption there exists an injective
mapping $\beta: V \rightarrow \Gamma$ that satisfies $\phi$.
Since $\Gamma$ is $l$-transitive,
 there is an automorphism $\gamma$ of $\Gamma$ that maps $\beta(a)$ to $b$
and that sends $\beta(w)$ to $\alpha(w)$, for $w \in T_i \setminus \{v\}$.
Then we extend $\alpha$ to $v$ by $\alpha(v) := \gamma(\beta(v))$; the extension
clearly satisfies $\phi$. Now we repeat the argument with $v$ in place of $a$, and $\alpha(v)$ in place of $b$, and are done by inductive assumption.
\end{proof}

\paragraph{\bf Application to the Rooted Phylogeny Problem.}
We now turn back to the rooted triple consistency problem, CSP$(\Lambda)$.
The structure $\Lambda$ is 2-transitive and the only relation has
arity three, and hence we can apply Lemma~\ref{lem:dom} to prove
that CSP$(\Lambda)$ cannot be solved by Datalog.

To construct an unsatisfiable girth $k$ instance $\Phi_k$ for CSP$(\Lambda)$,
let $G$ be a cubic graph of girth at least $k$ that has a Hamiltonian cycle.
Such a graph exists; see e.g.~the comments after the proof of Theorem 3.2 
in~\cite{BiggsCubic}. 
Note that $G$ must have an even number of vertices. 
Let $H = (v_1,v_2,\dots,v_n)$ be the Hamilton cycle of $G$.
For any vertex $a$ of $G$, let $r(a)$ be the vertex that precedes $a$ on $H$, $s(a)$ the vertex that follows $a$ on $H$, and $t(a)$ the third remaining neighbor of $a$ in $G$.

We now define $\Phi_k$. The vertices of $G$ will be the variables of $\Phi_k$. Then
$$ \Phi_k \; := \; \bigwedge_{a \in V(G)} r(a)s(a)|t(a) \; .$$
Consider the graph on the variables of $\Phi_k$ 
that has an edge $ab$
when $\Phi_k$ contains a triple clause $ab|c$ for some variable $c$ of $\Phi_k$. 
This graph is connected, since it actually equals the Hamilton cycle $H$ of $G$.
Hence, a condition due to Aho et al.~\cite{ASSU} implies that
$\Phi_k$ is unsatisfiable for all $k \geq 1$. This can also be seen by
Lemma~\ref{lem:rooted-phylo-unsat} in Section~\ref{sect:alg}. 
It is clear that every triple clause of $\Phi_k$ has an injective satisfying
assignment. 
So the only remaining condition to apply Lemma~\ref{lem:dom}
is the verification that $G(\Phi_k)$ has girth $k$.
But this is obvious since any cycle of length $2l < 2k$ in the incidence graph $G(\Phi_k)$ would give rise to
a cycle of length $l < k$ in $G$, in contradiction to $G$ having girth $k$. 

\begin{cor}
There is no Datalog program that solves the rooted triple consistency problem.
\end{cor}

\paragraph{\bf Other Applications of the Technique.}
Our technique to show Datalog inexpressibility
can be adapted to show that the following (closely related) problems cannot be
solved by Datalog as well.

\begin{enumerate}[$\bullet$]

    \item Satisfiability of branching time constraints~\cite{BroxvallJonsson};

    \item The network consistency problem of the left\hyp{}linear\hyp{}point
        algebra~\cite{Duentsch,HirschAlgebraicLogic};

    \item Cornell's tree description logic~\cite{Cornell,BodirskyKutzAI};

\end{enumerate}

All these three problems contain the following computational problem
as a special case.

\vspace{.2cm}
\begin{prob}[Tree-Description-Consistency]
\label{prob:pdc}
\cproblem{A finite structure $(V;<,||)$ where $<$ and $||$ are binary relations.}
         {Is there a rooted tree $T$ and $\alpha: V \rightarrow V(T)$
          such that if $x<y$ then $\alpha(y)$ lies strictly below $\alpha(x)$ in $T$,
          and if $x || y$ then neither $\alpha(x)$ lies below $\alpha(y)$ nor
          $\alpha(y)$ lies below $\alpha(x)$ in $T$?}
\end{prob}
\medskip

\noindent To again apply Lemma~\ref{lem:dom}, we first have to show that Tree-Description-Consistency
can be formulated as a CSP for a transitive 
$\omega$-categorical structure $\Omega= (D;<,||)$; this has already been observed in~\cite{BodirskyNesetrilJLC}. 
This time, it is more convenient to directly construct $\Omega$.
The domain $D$ consists of the set of all
non-empty finite sequences
of rational numbers. For 
$a = (q_1, q_2, \dots, q_{n}), b = (q'_1, q'_2, \dots, q'_{m})$, $n \leq m$, we write $a < b$ if one of the following conditions holds:
\begin{enumerate}[$\bullet$]
\item $a$ is a proper initial subsequence of $b$, i.e., $n<m$ and
$q_i=q_i'$ for $1 \leq i \leq n$;
\item $q_i=q_i'$ for $1 \leq i < n$, and $q_n < q_n'$.
\end{enumerate}
The relation $||$ is 
the set of all unordered pairs of distinct elements 
that are incomparable
with respect to $<$.
A proof that $\Omega$ is indeed $1$-transitive and $\omega$-categorical can be found in~\cite{AdelekeNeumann} (Section 5). Since the signature is binary, we can again apply Lemma~\ref{lem:dom}, and 
have to find unsatisfiable instances of arbitrarily high girth.

Here we use the fact that
Tree-Description-Consistency can simulate the rooted triple consistency problem by a simple reduction~\cite{BodirskyKutzAI}.
We construct $\Psi_k$ from $\Phi_k$
by replacing each triple clause of $\Phi_k$ of the form $xy|z$ by the three conjuncts
$u_{xyz} || z$, $u_{xyz} < x$, and $u_{xyz} < y$, where $u_{xyz}$
is a newly introduced variable. It can be shown (see~\cite{BodirskyKutzAI}) that this transformation preserves
(un-)satisfiability, and thus $\Psi_k$ is unsatisfiable as well. 
Moreover, the transformation is such that the girth of
$\Psi_k$ is not smaller than the girth of $\Phi_k$. 
Finally, it is clear that every conjunct in $\Psi_k$ has an injective satisfying
assignment.
Hence, Lemma~\ref{lem:dom} applies, and CSP$(\Omega)$
cannot be solved by Datalog.

\section{The Algorithm}\label{sect:alg}
In this section we show that the rooted phylogeny problem
can be solved in polynomial time if all clauses come from the
following class $\cal T$, defined as follows.

\begin{defi}\label{def:tame}
A disjunction $\psi := x_1 y_1|z_1 \lor \dots \lor x_p y_p|z_p$ is called
               \emph{tame} if it is trivial or 
               if $\{x_i,y_i\} = \{x_j,y_j\}$ for all $1 \leq i,j \leq p$.
	The set of all tame clauses is denoted by $\cal T$.
\end{defi}

The algorithm we present in this section builds
on previous algorithmic results
about the rooted triple consistency problem, most notably~\cite{ASSU,HenzingerKingWarnow}. 
One of the central ideas for the polynomial-time algorithm
for the rooted triple consistency problem in~\cite{ASSU}
is to associate a certain undirected graph to an instance of the
rooted triple consistency problem.
We generalize this idea to tame clauses as follows.

\begin{defi}
Let $\Phi$ be an instance of the rooted triple consistency problem
with tame clauses. Then
$F_\Phi:=(V,E)$ is the graph where the vertex set $V$ is the set of
variables of $\Phi$, and where $E$ contains an edge $\{x,y\}$ 
iff $\Phi$ contains a clause $x y|z_1 \lor \dots \lor x y|z_p$ for $p \geq 1$.
\end{defi}

The following provides a sufficient (but not a necessary) condition for unsatisfiability of rooted triple formulas with tame clauses.

\begin{lem}\label{lem:rooted-phylo-unsat}
Let $\Phi$ be an instance of the rooted phylogeny problem
with tame clauses. If $F_\Phi$ is connected then $\Phi$
is unsatisfiable.
\end{lem}
\begin{proof}
Let $V$ be the set of variables in $\Phi$.
Suppose that there is a solution $(T,\alpha)$ for $\Phi$.
Let $r$ be the yca of $\alpha(V)$ in $T$ (where $\alpha(V)$ is the set of
all leaves in the image of $V$ under $\alpha$).
It cannot be that all vertices in $\alpha(V)$ lie below the same child
of $r$ in $T$, since otherwise the child would have been above 
$r = \yca(\alpha(V))$, which is impossible. Since the graph $F_\Phi$ is connected, there is an edge $\{x,y\}$ in $F_\Phi$
such that $\alpha(x)$ and $\alpha(y)$ lie below different children of $r$ in $T$. Hence, there are $z_1,\dots,z_p \in V$ and a clause $x y|z_1 \lor \dots \lor x y|z_p$ in $\Phi$.
By assumption, the yca of $\alpha(x)$ and $\alpha(y)$, which is $r$, lies 
strictly below the yca of $\alpha(x)$ and $\alpha(z_i)$ for some $1 \leq i \leq p$, a contradiction
to the choice of $r$.
\end{proof}

To see that the condition is not necessary consider the following example.

\begin{exa}
	The rooted triple formula $\Phi = (ab|c \wedge bc|a \wedge ab|d)$ is unsatisfiable since the first two literals cannot simultaneously be satisfied. But the graph $F_{\Phi}$ is disconnected; it has the two components $\{a,b,c\}$ and $\{d\}$.
\end{exa}

\begin{figure*}[t]
\begin{center}
\fbox{
\begin{tabular}{l}
Solve$(\Phi)$ \\
Input: A rooted triple formula $\Phi$ with variables $V$ and clauses from $\cal T$. \\
Output: `yes' if $\Phi$ is satisfiable, `no' otherwise. \\
\\
If $\Phi$ is the empty conjunction then return `yes' \\
If $F_\Phi$ is connected \\
\hspace{.5cm} return `no' \\
else \\
\hspace{.5cm} Let $S$ be the vertices of a connected component of $F_\Phi$ \\
\hspace{.5cm} If Solve$(\Phi[S])$ is false or Solve$(\Phi[V \setminus S])$ is false return `no' \\
\hspace{.5cm} else return `yes' \\
\hspace{.5cm} end if \\
end if
\end{tabular}
}
\end{center}
\caption{The algorithm for the rooted phylogeny problem for tame clauses.}
\label{fig:assu}
\end{figure*}

\begin{thm}\label{thm:alg}
The algorithm {\it Solve} in Figure~\ref{fig:assu} determines
whether a given instance $\Phi$ of the rooted phylogeny problem for tame clauses is satisfiable.  When $m$ is the number of triples in all clauses, and $n$ is the number of variables of $\Phi$, then the algorithm can be implemented to run in time $O(m\log^2 n)$.
\end{thm}

\begin{proof}
If $\Phi$ is the empty conjunction, then $\Phi$ is clearly satisfiable,
and so the answer of the algorithm is correct in this case.
The algorithm first computes a
connected component $S$ of $F_\Phi$ (we discuss details of this step in the paragraph about the running time of the algorithm); 
if $S=V$, i.e., if $F_\Phi$ is connected,
then Lemma~\ref{lem:rooted-phylo-unsat} implies
that $\Phi$ is unsatisfiable. 

Otherwise, we execute the algorithm recursively on $\Phi[S]$ and on $\Phi[V\setminus S]$.
If any of these recursive calls
reports an inconsistency, then $\Phi$ is clearly
unsatisfiable as well: since if there was a solution $(T,\alpha)$ to $\Phi$, then $(T,\alpha|_V)$ would be a solution to $\Phi[V]$.
Otherwise, we inductively assume that the algorithm correctly
asserts the existence of a solution $(T_1,\alpha_1)$ of
$\Phi[S]$ and of a solution $(T_2,\alpha_2)$ of $\Phi[V\setminus S]$.

Let $T$ be the tree obtained by creating a new vertex $r$,
linking the roots of $T_1$ and $T_2$ below $r$, and
making $r$ the root of $T$.
Let $\alpha$ be the mapping that maps $x$ to $\alpha_i(x)$ if 
$x \in L(T_i)$, for $i \in \{1,2\}$.
We claim that $(T,\alpha)$ is a solution to $\Phi$,
i.e., we have to show that in every clause $\psi$ of $\Phi$ at least one
literal is satisfied.
If $\psi= (x y|z_1 \lor \dots \lor x y|z_p)$, 
then $x$ and $y$ are in the same subtree $T_i$ of $T$, since 
they are connected by an edge in $F_\Phi$. 
If all variables of $\psi$ lie completely inside $S$ or completely 
inside $V \setminus S$, we are done by inductive assumption,
because $(T_1, \alpha_1)$ is a solution for $\Phi[S]$
and $(T_2, \alpha_2)$ is a solution for $\Phi[V \setminus S]$.
Otherwise, there must be a $j$, $1 \leq j \leq p$, such that 
$z_j$ lies in a different component than $x$ and $y$.
But in this case the yca of $\alpha(x)$ and $\alpha(y)$ lies strictly below $r$, which is the yca of $\alpha(x)$ and $\alpha(z_j)$. Hence, the literal $xy|z_j$ in $\psi$ is satisfied.
This concludes the correctness proof of the algorithm shown in Figure~\ref{fig:assu}.

We still have to show how this procedure can be implemented
such that the running time is in $O(m \log^2 n)$. There are amortized sub linear algorithms for testing connectivity
in undirected graphs while removing the edges of the graph.
This was used to speed-up the algorithm for the rooted triple consistency problem~\cite{HenzingerKingWarnow}.
At present, the fastest known algorithm for this purpose appears to be 
the deterministic decremental graph connectivity
algorithm of Holm, de Lichtenberg, and Thorup~\cite{ThorupI}, which has a query time in $O(\log n/ \log\log n)$, and an update time in $O(\log^2 n)$.
We can use the same approach as in~\cite{HenzingerKingWarnow} and obtain an 
$O(m \log^2 n)$ bound for the worst-case running
time of our algorithm.
\end{proof}

\section{Complexity Classification}
\label{sect:hard}
This section is devoted to the proof of the following result.

\begin{thm}\label{thm:hard}
Let $\cal C$ be a set of rooted triple clauses that contains
clauses that are not tame (Definition~\ref{def:tame}).
Then the rooted phylogeny problem for clauses from
$\cal C$ is NP-complete.

\end{thm}

Our proof of Theorem~\ref{thm:hard} consists of two parts. 
In the first part, 
we show that if $\cal C$ is not a subset of $\cal T$, 
then a certain Boolean split problem associated to $\cal C$
(defined below) is NP-hard.
In the second part we show that this Boolean split problem 
reduces to the rooted phylogeny problem for $\cal C$.

\begin{defi}[split formula for $\Phi$]\label{def:split-problem}
Let $\Phi$ be a rooted triple formula. Then the \emph{split formula} for
$\Phi$ is the Boolean formula obtained from $\Phi$ by replacing 
each literal $xy|z$ by $(x \leftrightarrow y) \wedge (z \vee \neg z)$.
\end{defi}

The purpose of the tautological second conjunct $z \vee z$  is
to introduce the variable $z$, which would otherwise not appear 
in the formula; this becomes relevant in the following.
If $\cal C$ is a class of triple clauses,
we define $B(\cal C)$ to be the set of split formulas
for the clauses from $\cal C$. 

A solution to a propositional formula is called \emph{surjective} 
if at least one
variable is set to true and at least one variable is set to false.
The \emph{split problem} for a set of Boolean formulas $\cal B$ 
is the problem
to decide whether a given conjunction of formulas
obtained from formulas in $\cal B$ by variable substitution
has a surjective solution.

We will show that if $\cal C$ is a class of triple clauses that is not a
subclass of $\cal T$, then there exists a finite subset $\cal C'$ of $\cal C$
such that the split problem for $B(\cal C')$ is NP-complete. In the proof of
this statement we use the following result, which follows from Theorem 6.12
in~\cite{Creignou}, and is due to~\cite{CreignouHerbrard}. 
The notion of Horn,
dual Horn, affine, and bijunctive Boolean formulas are standard and introduced in detail in~\cite{Creignou}. Bijunctive formulas are also known as 2-CNF formulas.

\begin{thm}[of~\cite{CreignouHerbrard}]\label{thm:scsp-hard}
Let $\cal B$ be a set of Boolean formulas. Then the split problem for $\cal B$ is in P if all formulas in $\cal B$ are from one of the following types: Horn, dual Horn, affine, bijunctive. In all other cases, $\cal B$ contains a finite subset $\cal B'$ such that the split problem for $\cal B'$ is NP-complete.
\end{thm}

We say that a Boolean formula $\psi$ is \emph{preserved} by an operation $f:
\{0,1\}^k \rightarrow \{0,1\}$ if for all satisfying assignments
$\alpha_1,\dots,\alpha_k$ of $\psi$ the mapping defined by $x \mapsto
f(\alpha_1(x), \dots, $ $\alpha_k(x))$ is also a satisfying assignment for $\psi$.

\begin{prop}\label{prop:scsp-conditions}
If $\cal C$ is not a subclass of $\cal T$, then
$B(\cal C)$ is neither Horn, dual Horn, affine, nor bijunctive.
\end{prop}
\begin{proof}
Let $\phi$ be a clause from $\cal C \setminus \cal T$. By construction the split formula
$\psi$ for $\phi$ is preserved by $x \mapsto \neg x$ and is also preserved by
constant operations. 
Moreover, it is known (and follows from~\cite{Post}) 
that every Boolean formula that is
preserved by $\neg$, contains the constants, and is either
Horn, dual Horn, affine, or bijunctive must also be preserved
by the operation xor defined as $(x,y) \mapsto (x+y \mod 2)$.
So it suffices to show that $\psi$ cannot be preserved by xor.

Because $\phi$ is not from $\cal T$ and
in particular non-trivial,
there is a tree $T$ and an injective
mapping from the variables $V$ of $\phi$ to the leaves of $T$
such that $(T,\alpha)$ is not a solution to $\phi$.
Moreover, since the clause $\phi$ is not tame, it must contain triples $ab|c$ and
$uv|z$ where $\{a,b\} \neq \{u,v\}$.
Consider the assignment $\beta$ that maps $x \in V$ to $0$ if 
$\alpha(x)$ is below the first child of the yca of $\alpha(V)$ in $T$, and that maps $x$ to $1$ otherwise (which child is selected as
the first child is not important in the proof).
By construction, the assignment $\beta$ does not satisfy the split formula for
$\psi$, since $\phi$ is not satisfied by $(T,\alpha)$.
Observe that the assignment $\beta_1$
that is obtained from $\beta$ by negating the value
assigned to $a$ is a satisfying assignment for $\psi$, 
since it satisfies the disjunct $((a \leftrightarrow b) \wedge (c \vee \neg c))$ of $\psi$. The assignment $\beta_2$ 
that is constant $0$ except for the variable $a$ which is assigned $1$
is also a satisfying assignment for $\psi$, because $\psi$ satisfies
$((u \leftrightarrow v) \wedge (w \vee \neg w))$.
But since xor$(\beta_1(x),\beta_2(x))$ equals $\beta(x)$ for all $x \in V$,
this shows that $\psi$ is not preserved by xor, which is what we wanted to show.
\end{proof}

We now turn to the second part of the proof of Theorem~\ref{thm:hard}.
The idea to reduce the split problem for $B(\cal C)$ to 
the rooted phylogeny problem for clauses from $\cal C$ is to construct instances $\Phi$
of the phylogeny problem for $\cal C$ in such a way that $\Phi$
is satisfiable \emph{if and only if} $B(\Phi)$ has a 
surjective solution. To implement this idea, we construct an
instance of the phylogeny problem $\Phi$ that
fragments into simple and satisfiable pieces 
if $B(\Phi)$ has a surjective solution.

\begin{prop}\label{prop:reduction}
Let $\cal C$ be a finite class of triple clauses. Then
the split problem for $B(\cal C)$ can be reduced in polynomial time 
to the rooted phylogeny problem for clauses from $\cal C$.
\end{prop}

\begin{proof}
Note that the split formula for a trivial clause is a tautological Boolean formula. 
Hence, if all clauses in $\cal C$ are trivial, then the split problem for $B(\cal C)$ is clearly in P
and there is nothing to show. 
Otherwise, we can assume that
$\cal C$ contains the clause that just consists of $ab|c$ since this clause can be simulated by non-trivial clauses from $\cal C$ by appropriately equating variables (Lemma~\ref{lem:simtriple}).

Suppose we are given an instance of the split problem for $B(\cal C)$ 
with clauses $\psi_1,\dots,\psi_m$
and variables $V=\{x_0,\dots,x_{n-1}\}$.
We create an instance $\Phi$ of the rooted phylogeny problem for $\cal C$
as follows. 
The variables $U$ of $\Phi$ are triples $(x,i,j)$ where $x \in V$, $i \in \{0,\dots,m-1\}$, and $j \in \{1,\dots,n-1\}$. In the following, all indices of
variables from $V$ are modulo $n$. Moreover, if $m>1$ we will also
write $(x,i,n)$ for $(x,i+1,1)$ for all $i \in \{0,\dots,m-2\}$.
The clauses of $\Phi$ consist of two groups, $\Phi_1$ and $\Phi_2$. 

\begin{enumerate}[$\bullet$]
\item To define the first group $\Phi_1$ of clauses, suppose that $\psi_i$ has variables $y_1,\dots,y_q$.
Let $\phi_i(y_1,\dots,y_q)$ 
be the triple clause that defines
the Boolean relation from $B(\cal C)$ used 
in $\psi_i(y_1,\dots,y_q)$. 
By the assumption that 
$\cal C$ and $B(\cal C)$ are finite it is clear that $\phi_i$ can be computed efficiently (in constant time).
We then add the clause $\phi_i((y_1,i,1),\dots,(y_q,i,1))$
to $\Phi_1$. 
\item The second group $\Phi_2$ of clauses has for all $x_s \in V$, $i \in \{0,\dots,m-2\}$ (if $m=1$ the second group of clauses is empty), 
and $j \in \{1,\dots,n-1\}$ the
clause $$(x_s,i,j)(x_s,i,j+1)|(x_{s+j},i,1) \; .$$
\end{enumerate}
Note that $\Phi_2$ only consists of rooted triples, and therefore $F_{\Phi_2}$ is defined, and consists of exactly $n$ paths of length $(n-1)(m-1)$. 

We claim that $\Phi$ is satisfiable if and only
if $\psi_1 \wedge \dots \wedge \psi_m$ has a surjective solution.
First suppose that $\Phi$ has a solution $(T,\alpha)$.
Then the variables $U$ of $\Phi$ can be partitioned
into the variables that are mapped via $\alpha$ below
the left child of $\yca(\alpha(U))$, and the ones mapped below
the right child. Note that both parts of the partition are non-empty.
Variables $(x,i,j)$ of $U$ that share the first coordinate
are in the same part of the partition due to the clauses in $\Phi$ in the second group. Hence, the mapping that sends $x \in V$ to $0$
if $(x,i,j)$ is mapped to the first part, and that sends $x$ to $1$
otherwise is well-defined, and a surjective assignment.
It also satisfies all clauses $\psi_1,\dots,\psi_m$, 
because of the first group of
clauses in $\Phi$. 

Conversely, suppose that there is a surjective solution $s$ for $\psi_1 \wedge \dots \wedge \psi_m$. 
Let $S$ be the subset of the variables $V$ of $\Phi$ assigned to $0$ by $s$,
and consider the instances $\Phi_l := \Phi[S]$ and $\Phi_r := \Phi[V \setminus S]$. Since the assignment is surjective, there
is a variable $x_p \in V$ that is mapped to 1 and a variable $x_q \in V$ that is mapped to $0$. Hence, for all $i \in \{0,\dots,m-1\}$ 
the clauses $(x_p,i,q-p)(x_p,i,q-p+1)|(x_q,i,1)$ from the second group are neither in $\Phi_l$ nor in $\Phi_r$,
because they contain variables from both parts of the partition.
Therefore, any clause from the first group in $\Phi_l$ 
will be disconnected in the incidence graph $G(\Phi_l)$ from any other clause in the first group in $\Phi_l$. Since each clause from the first group is satisfiable, 
it is easy to see that $\Phi_l$ has a solution
$(T_l,\alpha_l)$. 
The same statements holds for $\Phi_r$; let $(T_r,\alpha_r)$
be a solution for $\Phi_r$. Let $T$ be the rooted tree obtained
from $T_l$ and $T_r$ by creating a new vertex $t$,
linking the roots of
$T_l$ and $T_r$ below $t$, and making $t$ the root of $T$.
Let $\alpha$ be the common extension of $\alpha_l$ and $\alpha_r$ to all of $U$. Then $(T,\alpha)$ is clearly a solution
to $\Phi$. 

Both groups of clauses together consist of $m+n(n-1)(m-1)$ many clauses,
and it is easy to see that the reduction can be implemented in
polynomial time.
\end{proof}

We conclude this section with a combination of the results above.

\begin{proof}[Proof of Theorem~\ref{thm:hard}]
As mentioned, the rooted phylogeny problem
for $\cal C$ is clearly in NP. 
Let $\cal C$ be a class of triple clauses that is not
a subset of $\cal T$. We prove NP-hardness as follows.
By Proposition~\ref{prop:scsp-conditions}, 
$B(\cal C)$ is neither Horn, dual Horn, affine, nor bijunctive.
Theorem~\ref{thm:scsp-hard} asserts that there exists a finite
subset $\cal B$ of $B(\cal C)$ such that the split problem for $\cal B$ is NP-hard.
This means that there is a subset $\cal C'$ of $\cal C$ such that
the split problem for $B(\cal C')$ is NP-hard. Proposition~\ref{prop:reduction}
shows that the rooted phylogeny problem for clauses from $\cal C'$
(and hence also for clauses from $\cal C$) is NP-hard as well.
\end{proof}

\section{Concluding Remarks}

We have shown that consistency of rooted phylogeny data 
can be decided in polynomial time when
the data consists of tame disjunctions of rooted triples. 
Our algorithm extends previous algorithmic results about the rooted triple
consistency problem, without sacrificing worst-case efficiency. 
The class $\cal T$ of
tame triple clauses that can be handled efficiently is also
motivated by another result of this paper, which states that any set of 
triple clauses that is not contained in $\cal T$ has an NP-complete rooted
phylogeny problem. Here we use known results about the complexity of surjective Boolean constraint satisfaction problems. 

We also show that no Datalog program can solve the rooted
triple consistency problem, using a pebble game that
captures the expressive power of Datalog for constraint satisfaction problems with infinite $\omega$-categorical structures.
In fact, our result follows from a more general result 
that also applies to many
constraint satisfaction problems outside of phylogenetic reconstruction. We show that a constraint satisfaction problem
for a structure with a large automorphism group
cannot be solved by Datalog if, roughly, for all $k$ there exists a unsatisfiable instance of girth at least $k$.

The class of phylogeny problems studied in this paper 
has a natural generalization to a larger class of computational problems, namely problems of the form 
CSP$(\Gamma)$ where $\Gamma$ has a first-order definition in $\Lambda$, the $\omega$-categorical relatively 3-transitive C-set introduced in Section~\ref{sect:datalog}.
This class contains several additional problems that have been studied in phylogenetic reconstruction, for instance the quartet consistency problem~\cite{Steel}. 
The larger class also contains new problems that can be solved in polynomial
time, and where the split problem consists in finding surjective
solutions to Boolean linear equation systems.
A complexity classification for this larger class of computational
problems remains open and is left for future research.

\section*{Acknowledgement}

We would like to thank the reviewers for their helpful comments.

\bibliographystyle{alpha}
\bibliography{global}

\end{document}